\documentclass[times, twoside]{zHenriquesLab-StyleBioRxiv-Submission}
\usepackage{blindtext}
\usepackage{graphicx}
\usepackage{dblfloatfix} 
\usepackage{comment}
\usepackage[normalem]{ulem}
\usepackage{color, colortbl}
\usepackage{MnSymbol}
\usepackage{amsmath}
\usepackage[mathscr]{euscript}

\renewcommand{\thesubsection}{\thesection\Alph{subsection}}

\titleformat{\subsection}
  {\sffamily\bfseries\itshape}
  {\thesubsection.}
  {0.5em}
  {#1}
  []
\renewcommand{\footerfont}{\normalfont\sffamily\fontsize{8}{9}\selectfont}

\leadauthor{Madabhushi Balaji} 

\begin{document}

\title{Lensless wide-field 3D fiber endoscopy \\ through scattering media using \\ synthetic wavelength holography}


\author[1,3,*]{Muralidhar Madabhushi Balaji}
\author[1,3]{Parker Liu}
\author[1]{Patrick Cornwall}
\author[1]{Tianyi Wang}
\author[1,2]{Juergen Czarske}
\author[1,*]{ \\ Florian Willomitzer}

\affil[1]{Wyant College of Optical Sciences, University of Arizona, Tucson, AZ, 85721, USA.}
\affil[2]{Department of Electrical \& Computer Engineering, TU Dresden, Germany.}
\affil[3] {These authors contributed equally to this work.}
\affil[*] {mmuralidhar@arizona.edu, fwillomitzer@arizona.edu}

\maketitle

\vspace{-3mm}

\section*{Abstract} 
\textit{Minimally invasive imaging with fiber optic endoscopes is crucial for in vivo visualization of tissue morphology, as it supports applications such as early detection of tumors. However, imaging performance of conventional fiber endoscopes is limited when scattering layers are present between target and the distal end of endoscope. This limitation is particularly relevant in biomedicine, where targets such as early stage lesions or blood clots can be obscured by scattering tissue. To address this challenge, we present a lensless endoscopic imaging approach based on synthetic wavelength holography (SWH). SWH is a computational imaging technique in which two optical fields acquired at closely spaced wavelengths are combined to synthesize a field at a longer synthetic wavelength. As the field at longer wavelengths is less sensitive to path length perturbations, this approach can enable endoscopic recovery of holographic information despite scattering in the intervening tissue. In addition, because the synthetic field is assembled from scattered optical fields with larger optical etendue, our approach can extend the field of view (FoV) beyond the numerical aperture of the fiber. In this paper, we present the first demonstration of an SWH-based lensless endoscope using a multicore multimode fiber. We experimentally recover three-dimensional images of objects hidden behind scattering layers and through real biological tissue, with a spatial resolution of $\approx 500~\mu m$. We further demonstrate recovery of object information over an extended FoV of $46^\circ$ without additional distal optics. These results suggest a practical path toward extending fiber endoscopy for wide-field, three-dimensional imaging through scattering media.}

\section{Introduction}
\noindent Fiber endoscopes have proven to be powerful and versatile tools for a wide range of imaging and sensing applications in biomedicine~\cite{sun_lensless_2024}. Among the different fiber platforms, multi core optical fibers (MCFs) are particularly attractive because they combine a compact and flexible form factor with high spatial sampling density. These properties have enabled a variety of biomedical imaging architectures in recent years~\cite{potter2024clinical,conkey2016lensless,bae2023feasibility, gau2024multicore,choi2022flexible,du2022hybrid}. For example, MCF-based probes have been used to develop lensless multiphoton endoscopes for high-resolution cellular imaging~\cite{conkey2016lensless,bae2023feasibility}, they enable minimally invasive probing of brain activity~\cite{gau2024multicore}, and image unstained biological tissues~\cite{choi2022flexible}. In addition, the high core density of MCFs has enabled lensless endoscopes that reconstruct object images from holographic measurements~\cite{sun_quantitative_2022,badt_real-time_2022,sun_lensless_2024,scharf2019holographic,dremel2026lensless}. These platforms leverage digital holographic principles to bypass the need for distal optics, thereby further reducing the footprint of the endoscope.

Beyond biomedical applications, the same properties that make MCFs attractive for endoscopy have also been harnessed for broader applications in optical sensing, communications and computational imaging~\cite{richardson2013space,porat_widefield_2016,french_snapshot_2018}. For instance, MCFs have been utilized to realize compact hyperspectral imagers by capturing complex-valued optical fields across multiple wavelengths \cite{french_snapshot_2018}. Likewise, MCF endoscopes are also used to recover the topographic measurements of technical parts using multi wavelength interferometry \cite{groger2024two}. Furthermore, angular correlations in the speckle patterns emerging from the MCFs have been used for developing wide field lensless imaging platforms~\cite{porat_widefield_2016}.

\begin{figure*}[ht!]
    \centering
    \includegraphics[width=0.9\linewidth]{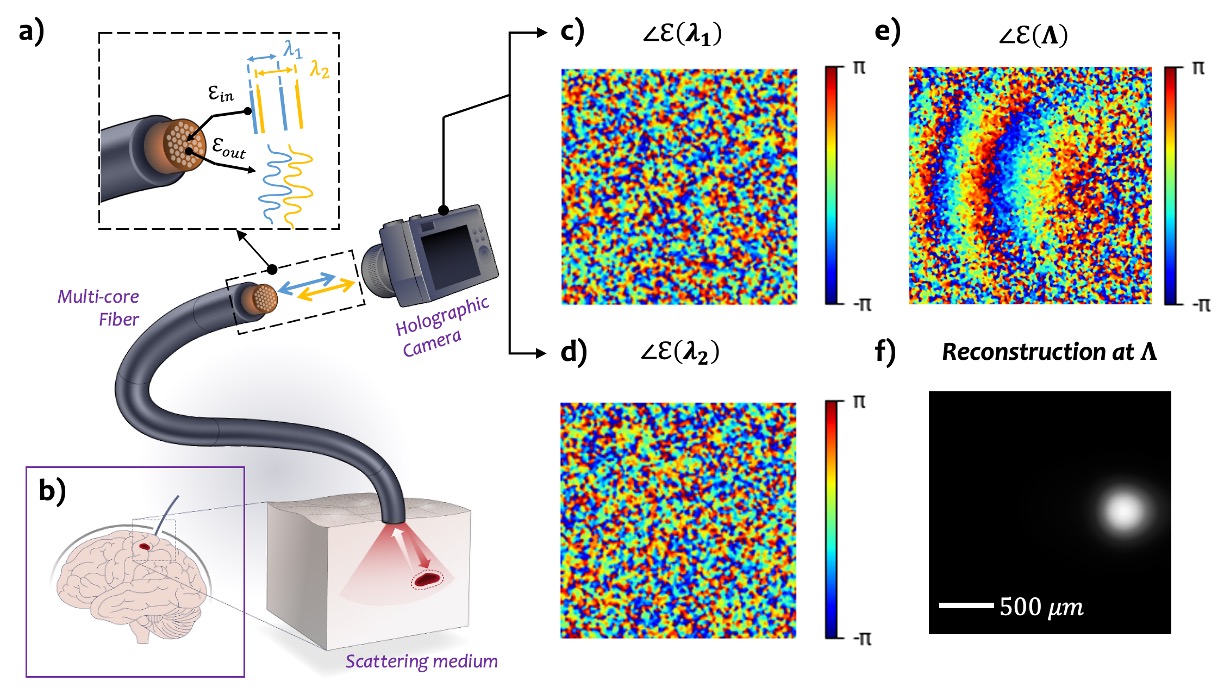}
    \caption{\textbf{Lensless wide-field 3D endoscopy through scattering media using synthetic wavelength holography.} a) \textit{Conceptual overview of the imaging configuration.} Light from two lasers operating at closely spaced wavelengths, $\lambda_1$ and $\lambda_2$, is coupled into a multicore fiber positioned in direct contact with the tissue surface. The resulting scattered optical field propagates through the tissue and interacts with an embedded object, such as a tumor. The optical fields emerging at the tissue surface are collected by the multimode multicore fiber and relayed to the holographic camera (methods), where the corresponding holograms are recorded. b) \textit{Illustration of a potential application.} Conceptual depiction of the proposed approach in minimally invasive procedures involving brain tissue. c) The phase maps of the speckle fields recorded by the holographic camera at $\lambda_1$ (for details, see Supplementary Sec. 2). d) The phase map of the speckle fields recorded by the holographic camera at $\lambda_2$ e) The phase map of the field at the synthetic wavelength $\Lambda$ f) Reconstruction of the point source  via holographic backpropagation of the captured synthetic field at $\Lambda$.}
    \label{fig:fig1}
\end{figure*}

However, despite these developments, both MCF-based and non MCF-based fiber endoscopes are primarily limited to regimes in which the object remains in the direct line of sight of the distal fiber tip. This constraint reduces their effectiveness in scenarios that require imaging through scattering environments, such as resolving morphological features beneath tissue layers. When imaging through biological tissue layers, the incident light is significantly scattered, causing a degradation in the detected imagery. This limitation has motivated extensive research on imaging through scattering media in recent times \cite{bertolotti2022imaging, yoon2020deep, bertolotti2012non, katz2014non, edrei2016optical, rangarajan2019spatially, lindell2020three, feng2023neuws, balaji2024probing, kassem2025intensity}. 

Additionally, MCF-based endoscopes are constrained by the narrow instantaneous angular field of view (FoV), due to the finite acceptance angle of the individual fiber cores. In practice, a limited field of view can reduce spatial context, making it more difficult to localize anatomical features of interest, or support surgical guidance ~\cite{perperidis2020image}. In coherent imaging platforms, the use of MCFs also introduces additional challenges, such as high sensitivity to physical and environmental perturbations such as fiber bending, surface height irregularities at the fiber facets \cite{sun_quantitative_2022}. Moreover, even MCFs comprising single-mode cores also exhibit multimode behavior due to inter-core crosstalk, leading to scrambling of the transmitted phase.

In this manuscript, we address the above limitations by developing a robust approach that enables endoscopic imaging over a wide FoV through scattering media, such as real biological tissue. Our approach is based on the emerging computational imaging technique called \textit{Synthetic Wavelength Holography} (SWH)~\cite{willomitzer_fast_2021,willomitzer2019synthetic,willomitzer2024synthetic, forschner2024towards, balaji2025fiber, liu20263d}.   In SWH, we illuminate the scattering medium using temporally coherent light at two closely spaced optical wavelengths $\lambda_{1}$ and $\lambda_{2}$, either directly through the MCF or through an external illumination fiber. The scattered light from the medium interacts with the object, and propagates back through scattering medium and MCF to reach the sensor. The optical fields at these two wavelengths, $\mathscr{E}(\lambda_1)$ and $\mathscr{E}(\lambda_2)$, are then recorded at the sensor. By \textit{computationally} mixing the two optical fields, $\mathscr{E}(\lambda_1)$ and $\mathscr{E}(\lambda_2)$, we assemble a complex-valued field representation $\mathscr{E}(\Lambda)$ that can be viewed as a hologram of the object at a \textit{synthetic wavelength}, $\Lambda  = \frac{\lambda_{1}\lambda_{2}}{|\lambda_{1} - \lambda_{2}|}$ (see Sec. 2.1). 

The object information within the two optical fields, $\mathscr{E}(\lambda_1)$ and $\mathscr{E}(\lambda_2)$, is scrambled by the phase perturbations introduced through modal scrambling in the MCF and the scattering medium. But the field at the longer synthetic wavelength, $\Lambda$, is less sensitive to the phase perturbations introduced by scattering, enabling the recovery of the hologram of the object (see Supplementary Sec. 1 for further details). This reduced sensitivity can be understood intuitively by noting that the \textit{computationally} assembled field, $\mathscr{E}(\Lambda)$, largely behaves as if it were generated by a physical electromagnetic wave at the longer wavelength~$\Lambda$, as demonstrated in \cite{willomitzer_fast_2021, willomitzer2024synthetic}.

Furthermore, the generation of the field at the synthetic wavelength from optical carriers offers several additional advantages. First, despite effectively measuring the object field at longer effective wavelengths (e.g., in the THz regime), the intrinsic contrast from the optical illumination is preserved in our approach. In addition, since the synthetic wavelength $\Lambda$ is predominantly determined by the difference between $\lambda_{1}$ and $\lambda_{2}$, it can be flexibly tuned over orders of magnitude by adjusting the illumination wavelength of one of the lasers. Finally, the fact that our carrier is light at optical wavelengths allows us to perform all the experiments shown in this paper using standard tunable continuous-wave (CW) lasers and off-the-shelf CMOS sensors, thereby eliminating the need for dedicated THz or long wave infrared (LWIR) sources and  detectors.

The advantages above also extend to the MCFs used in this work. Specifically, they allow an MCF designed for optical wavelengths to be effectively repurposed as a THz MCF, eliminating the need for using fibers that are operable at THz wavelengths. Furthermore, MCFs composed of multimode cores support multiple spatial modes at optical wavelengths, leading to the formation of speckle. Consequently, the phase of light transmitted through such multi-core, multi-mode fibers (M\textsuperscript{3}CFs) is randomized at optical wavelengths. However, at the longer synthetic wavelength, these cores support only a single spatial mode, preserving the transmitted phase information at the synthetic wavelength. Thus, the synthetic-wavelength hologram captured at the distal end of the fiber can be used to reconstruct the obscured object through computational back-propagation.

Consequently, the proposed approach has the potential to expand the use of fiber optic endoscopes for demanding applications such as imaging through brain tissue, see Fig.~\ref{fig:fig1}a. In this scenario, two lasers at closely spaced wavelengths can be used to illuminate the scattering tissue. The resulting scattered speckle fields, which carry information from the object, propagate back through a multicore fiber and are recorded using a single-shot holographic camera (described in Methods). Experimental results from the proposed SWH based endoscope are shown in Fig.~\ref{fig:fig1}c--f. Here, we use a multicore multimode fiber (M\textsuperscript{3}CF) based endoscope to record the synthetic wavelength hologram of an an off-axis point source hidden behind a ground glass diffuser (see Supplementary Sec.~2 for details of the experimental setup). We can see from  Figs.~\ref{fig:fig1}c,d that the phase maps of the acquired optical holograms, $\angle \mathscr{E}(\lambda_1)$ and $\angle \mathscr{E}(\lambda_2)$, are randomized by the combination of scattering at the diffuser and multimode propagation through the M\textsuperscript{3}CF.

However, the phase map of the computationally assembled synthetic wavelength hologram, $\angle \mathscr{E}(\Lambda)$, exhibits iso-contours corresponding to a spherical wavefront emanating from the off-axis point-source object, as shown in Fig.~\ref{fig:fig1}e. To assemble $\mathscr{E}(\Lambda)$, we compute the pixel-wise complex-conjugate product of the two normalized optical fields, $\mathscr{E}(\Lambda) = {\mathscr{E}(\lambda_1)\mathscr{E}^{*}(\lambda_2)}$. The assembled field, $\mathscr{E}(\Lambda)$, is then computationally backpropagated at the synthetic wavelength to reconstruct the off-axis point source, as shown in Fig.~\ref{fig:fig1}f.

\vspace{2mm}

We summarize the main contributions of this work and the prospective application areas below: 

\begin{itemize}
    \item \textbf{Lensless endoscopic imaging through scattering media:} We introduce an SWH-based lensless fiber endoscope capable of imaging objects obscured by a scattering medium. We show that this approach is robust to modal scrambling within the fiber and scattering in the intervening tissue by demonstrating its effectiveness in transmission and reflection geometries, as well as through real biological tissue. These results underscore the potential of this technology to enable endoscopic detection of subsurface features, such as vasculature or tumors beneath human brain tissue (see Fig.~\ref{fig:fig1}).

    \item \textbf{Volumetric 3D localization of objects obscured by scattering media:} We leverage holographic phase information for three-dimensional localization of objects hidden behind scattering media. This is accomplished through numerical backpropagation of holograms captured at multiple synthetic wavelengths. This capability could potentially support applications like volumetric mapping of tumors beneath tissue layers.

    \item \textbf{Field-of-View (FoV) expansion:} The angular FoV of conventional lensless fiber endoscopes is typically limited by the numerical aperture ($NA$) of the fiber cores. However, the presence of a scattering medium between the distal end of the fiber and the object increases the étendue of the imaging system, enabling light collection over a wider FoV. However, the scattered light from the wider FoV is typically scrambled and unusable. We use SWH to recover the object information encoded in the scattered light, and thereby image over a substantially increased angular FoV that can, in principle, span the entire hemisphere. This capability could potentially allow surgeons to visualize a wider surgical volume without requiring mechanical scanning or repositioning the endoscope.

    \item \textbf{Improved robustness to fiber bending:} Our experiments demonstrate the robustness of the SWH endoscope to dynamic external perturbations, such as fiber bending. This is due to the fact that the synthetic wavelength is orders of magnitude longer than the optical wavelength. In clinical settings, this resilience could simplify endoscope handling and reduce the need for complex, real-time recalibration when the endoscope moves during a procedure.
\end{itemize}

Although the larger synthetic wavelength improves robustness to phase perturbations and scattering, it does so at the cost of spatial resolution  of the object reconstructions. Consequently, the spatial resolution of our approach will be lower than that of optical approaches at single wavelength. However, with appropriate choices of fibers and tunable lasers (see Discussion section), we anticipate to further improve our currently demonstrated  $\approx 500~\mu m$ resolution  towards resolutions in the $50~\mu m$ to $100~\mu m$ range, which would make the approach competitive for applications such as optogenetics \cite{accanto2023flexible}. In the following sections, we present experimental results that demonstrate the performance of our technique, examine the respective performance limits and trade-offs, and discuss the broader implications and methodological details of the approach.

\section{Results}

\subsection{Measurement technique and experimental setup}
\label{subsec:expsetup}
\begin{figure}[t]
    \centering
    \includegraphics[width=\linewidth]{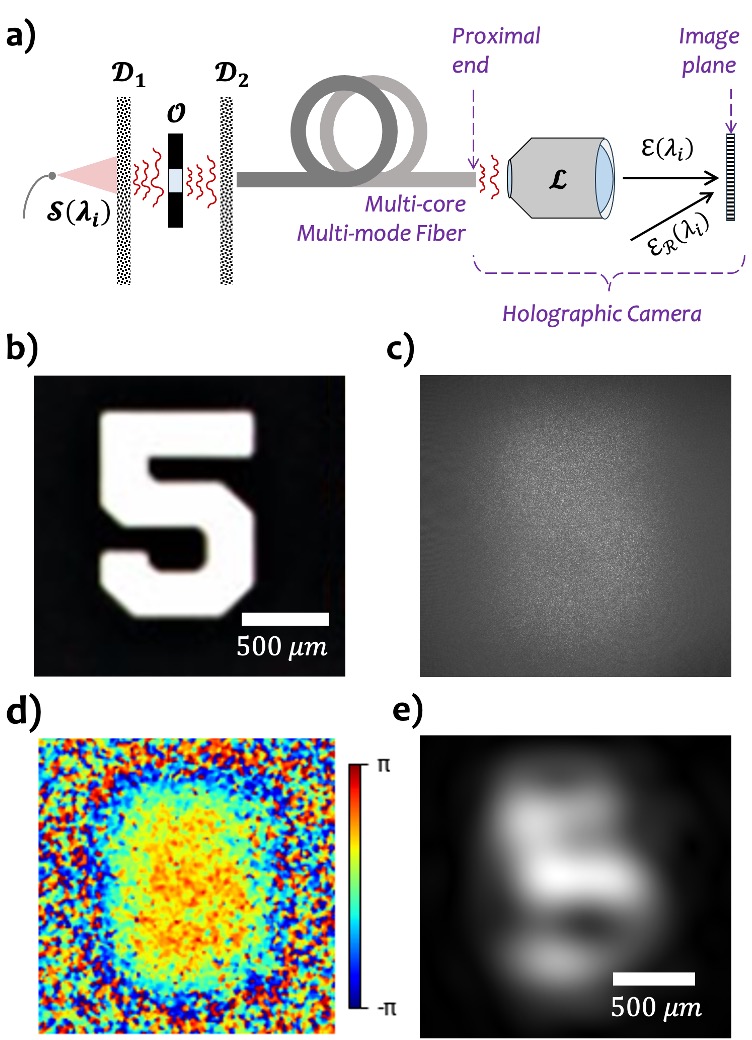}
    \caption{\textbf{Lensless endoscopy through discrete scattering layers in transmission geometry.} a) Schematic illustration of the setup. A USAF test target embedded between two ground glass diffusers $\mathscr{D}_{1}$ and $\mathscr{D}_{2}$ is imaged using our method. b) Image of the embedded object c) Intensity image observed through the endoscope d) Phase map at the synthetic wavelength $\Lambda = 500~\mu m$ obtained using our approach and e) Reconstruction of the embedded object obtained by back propagating the field at the synthetic wavelength.}
    \label{fig:fig2}
\end{figure}

\noindent We use the endoscopic system schematically depicted in Fig.~\ref{fig:fig2}a to acquire synthetic wavelength holograms through a scattering medium. The setup uses two narrow-linewidth lasers with emission wavelengths $\lambda_1$ and $\lambda_2 = \lambda_1 + \Delta\lambda$, separated by a predefined wavelength offset $\Delta\lambda$. We positioned a M\textsuperscript{3}CF ($3~mm$ diameter comprising of $3012$ individual cores of diameter $50\,\mu\text{m}$) in direct contact with the scattering medium to collect the light emerging from the scattering medium. Although each core within this fiber is multi-mode at optical wavelengths and introduces additional scrambling of the propagating light, the field at the larger synthetic wavelength exhibits robustness to such perturbations (Supplementary Sec.1). In the proof-of-principle experiments shown in this manuscript, we use a separate single-mode fiber to illuminate the scattering medium and deliver light to the target, either in a transmission setting as shown in Fig.~\ref{fig:fig2}a, or in reflection as shown in Figs. \ref{fig:fig3}a, \ref{fig:fig4}a, and \ref{fig:fig5}a. This architecture can be readily adapted to directly illuminate the scattering medium through the M\textsuperscript{3}CF, thereby facilitating a truly minimally invasive configuration.

The light incident on the scattering medium is scattered towards the object and illuminating it with a speckle pattern. The light from the object is scattered back towards the interface between the scattering medium and the endoscope, where it is collected by the M\textsuperscript{3}CF.  This light distribution at the proximal end of the M\textsuperscript{3}CF is imaged onto a CMOS sensor, where the two optical fields $\mathscr{E}(\lambda_1)$ and $\mathscr{E}(\lambda_2)$ are recorded. To record these optical fields, we rely on an an off-axis holographic configuration, where the reference beams, $\mathscr{E}_{R}(\lambda_i)$, are launched at an angle towards the CMOS sensor. This sub-system is detailed in \textit{Methods} and referred to as ``holographic camera'' in this paper. The two optical fields can be acquired either sequentially or in a single shot using a spatial-division multiplexing scheme developed in our prior work~\cite{ballester_single-shot_2024}. The synthetic wavelength hologram is then assembled as described earlier, by computing the per-pixel complex-conjugate product of these two optical fields, i.e., $\mathscr{E}(\Lambda)= \mathscr{E}(\lambda_{1})\cdot \mathscr{E}^{*}(\lambda_{2})$. The three-dimensional reconstruction of the object is then obtained by backpropagating the synthetic-wavelength field, $\mathscr{E}(\Lambda)$, using the angular-spectrum backpropagation algorithm.

\subsection{Lensless 3D endoscopy through thin scatterers}
\label{subsec:thinscats}

\noindent We use the apparatus shown in Fig.~\ref{fig:fig2}a to demonstrate the operation of our SWH based lensless endoscopy through  discrete scattering layers in transmission geometry. In this experiment, we recover the holographic information of a USAF target, $\mathscr{O}$, embedded between two ground glass diffusers, $\mathscr{D}_{1}$ and $\mathscr{D}_{2}$. Here, a probe beam from the tunable lasers is incident on the first ground glass diffuser ($\mathscr{D}_{1}$), which scatters the light and illuminates a small region \textit{(Number $5$ in Group $0$)} on the transmissive USAF target (see Fig.~\ref{fig:fig2}b). The light from the USAF target then propagates through free space to reach the second ground glass diffuser ($\mathscr{D}_{2}$), where the M\textsuperscript{3}CF placed in close proximity to $\mathscr{D}_{2}$ collects the light. The optical field at the proximal end of the M\textsuperscript{3}CF is then recorded by the holographic camera. 

We can see from Fig.~\ref{fig:fig2}c that the recorded intensity image of the object is completely scattered and does not display any prominent features of the illuminated object. This is due to a combination of scattering events at the two diffusers $\mathscr{D}_{1}$ and $\mathscr{D}_{2}$, and phase scrambling in the multimode cores of the M\textsuperscript{3}CF. However, the synthetic wavelength hologram, assembled at a synthetic wavelength of $\Lambda = 500 \mu m$, retains the phase information of the obscured object at the synthetic wavelength, as shown in Fig~\ref{fig:fig2}d. We backpropagate this synthetic wavelength hologram to reconstruct the intensity image of the obscured object (see Fig.~\ref{fig:fig2}e). The reconstructed image clearly resolves the features of the number 5 in Group 0, at an approximate spatial resolution of $500 \mu\text{m}$ (see Supplementary Sec. 2). 

Although smaller $\Lambda$ can potentially improve the resolution, the reconstructions become progressively noisy due to the spectral decorrelation between the field at the two wavelengths. This effect can be seen in the previous demonstrations of synthetic wavelength imaging  \cite{willomitzer2019synthetic, willomitzer_fast_2021}. For the configurations shown in this manuscript, we have evaluated $500~\mu m$ as the optimal $\Lambda$ that balances resolution and robustness to scatter. We refer to the \textit{Discussion} section for further details on this topic.

\begin{figure}[t]
    \centering
    \includegraphics[width=\linewidth]{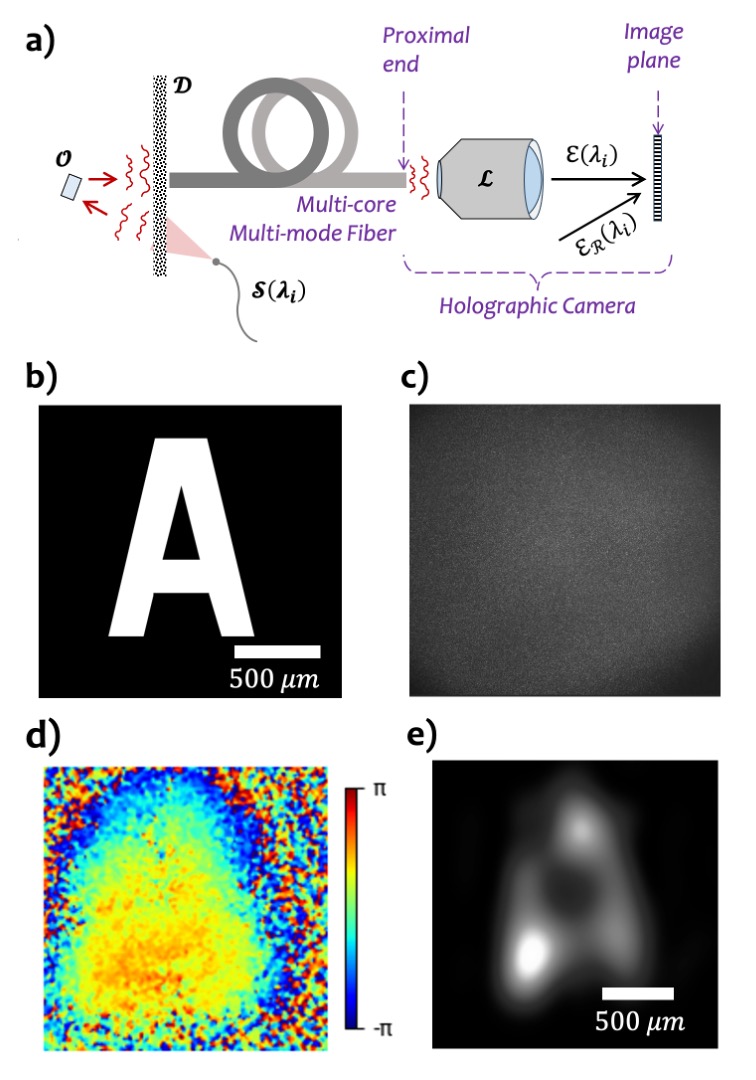}
    \caption{\textbf{Lensless endoscopy through discrete scattering layers in reflection geometry.} a) Schematic illustration of the setup. A reflective test target is placed behind a ground glass diffuser $\mathscr{D}$ is imaged using our method. b) Image of the obscured object c) Intensity image observed through the endoscope d) Phase map at the synthetic wavelength $\Lambda = 500~\mu m$ obtained using our approach and e) Reconstruction of the obscured object obtained by back propagating the field at the synthetic wavelength.}
    \label{fig:fig3}
\end{figure}

We also performed experiments in  reflection geometry using the setup schematic shown in Fig.~\ref{fig:fig3}a. Here, we recover a synthetic wavelength hologram of a reflective object $\mathscr{O}$ placed behind a ground glass diffuser using the same M\textsuperscript{3}CF based endoscope as before. The probe beam from the tunable lasers is incident on the ground glass diffuser ($\mathscr{D}$), which scatters the light and illuminates the object \textit{(mask of letter "A" on a mirror)}. The scattered light is reflected by the object to reach the diffuser $\mathscr{D}$, where it is again scattered and coupled into the  M\textsuperscript{3}CF. We then record the hologram of the light distribution at the proximal end of the fiber using the holographic camera.

Similar to the previous experiment, we can see from Fig.~\ref{fig:fig3}c that the object behind the scattering layer is completely obscured to the holographic camera due to scattering, but the phase distribution of the field at the synthetic wavelength of $\Lambda = 500~\mu m$, is not randomized and exhibits structure (see Fig~\ref{fig:fig3}d). We backpropagate the respective field at the synthetic wavelength to reconstruct the obscured object. From the reconstruction shown in Fig.~\ref{fig:fig3}e, we see that the shape of the object can be recovered with an approximate spatial resolution of $500 \mu\text{m}$ 

\begin{figure}[t!]
    \centering
    \includegraphics[width=\linewidth]{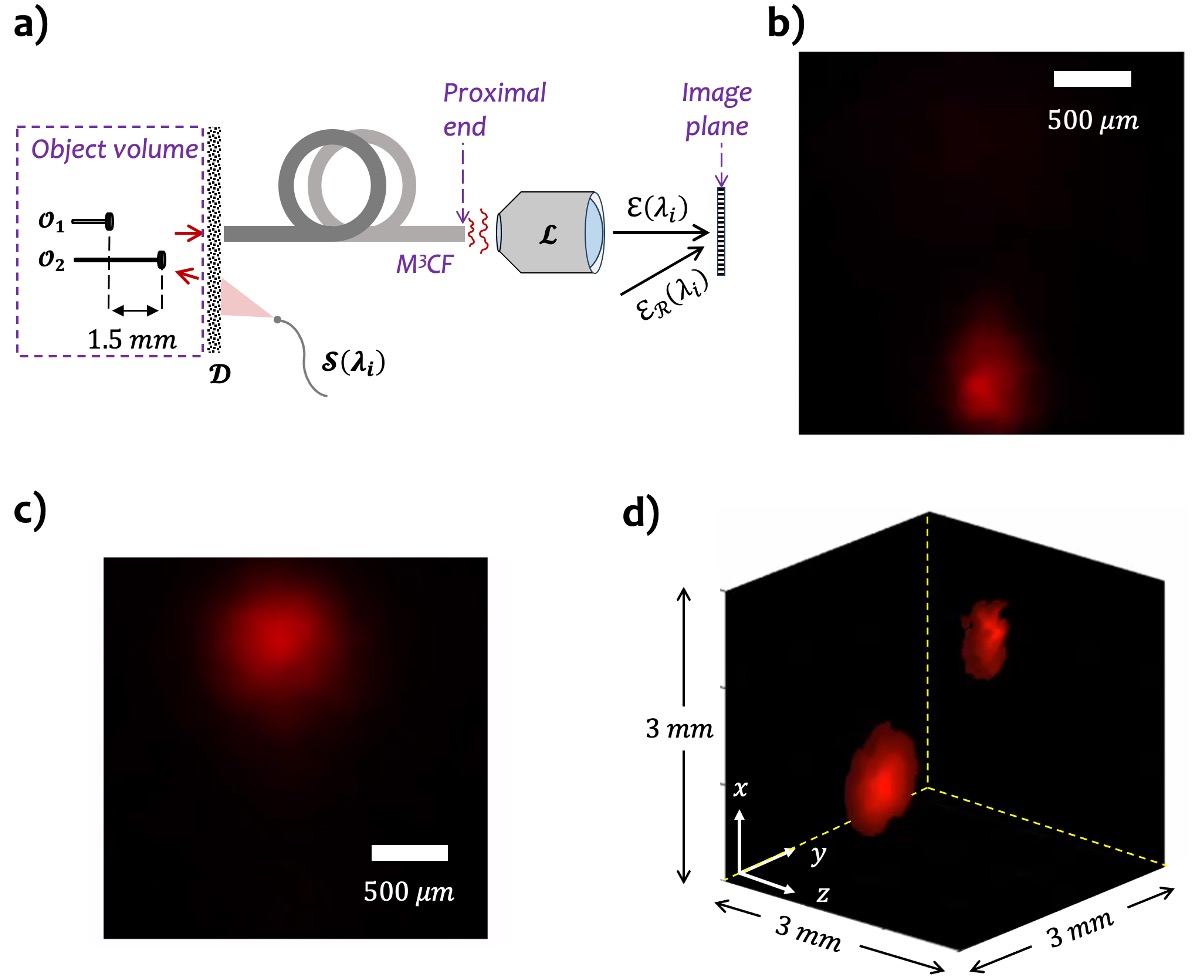}
    \caption{\textbf{Localizing objects behind a scattering medium.} a) Schematic  of the experimental setup. Two reflective pins, $\mathscr{O}_1$ and $\mathscr{O}_2$ were placed at two distinct axial positions separated by $1.5~mm$ behind a ground glass diffuser. b) Reconstructed image at $1~mm$ depth c) Reconstructed image at $2.5~mm$ depth d) Volumetric reconstruction.}
    \label{fig:fig4}
\end{figure}

In the experiment shown in Fig.~\ref{fig:fig4}, we used the same endoscope system to demonstrate volumetric 3D localization of obscured objects behind scattering layers. We do this by computational superposition of a sequence of synthetic wavelength holograms recorded at uniformly spaced, linearly increasing synthetic wavelengths \cite{cornwall2024synthetic}. This computational superposition operation results in a synthetic pulse wavefront, resembling the process of realizing a frequency comb, albeit through computational means (Methods). The assembled synthetic pulse is then computationally advanced or delayed to reconstruct the three-dimensional volume surrounding the object. We refer to Methods and \cite{cornwall2024synthetic} for further details.

In our experiments, we measured two reflective pins, $\mathscr{O}_1$ and $\mathscr{O}_2$ placed at two distinct axial positions separated by $1.5~mm$ in the volume behind a ground glass diffuser, $\mathscr{D}$ (see Fig.~\ref{fig:fig4}a). We used our computational synthetic pulse assembly to reconstruct the three-dimensional volume of the obscured objects at a calculated depth resolution of $500~\mu m$. In Fig.~\ref{fig:fig4}b and c, we show the volumetric reconstructions of the  pins at $1~mm$ and $2.5~mm$ distance to the scatterer. It can be seen that the objects are distinctly separated in the axial dimension and no spurious artifacts from the other object are present. 

The obtained reconstructions can also be rendered as 3D volume by stacking all the reconstructed depth planes. In Figure \ref{fig:fig4}d, we display a reconstruction of a $3~\text{mm} \times 3~\text{mm} \times 3~\text{mm}$  volume hidden behind the scattering layer using our approach. This capability to acquire volumetric reconstructions through a thin scattering layer, could extend minimally invasive fiber endoscopy beyond surface or 2D imaging toward volumetric mapping of tissue microstructure without requiring distal optics. Such an approach could support applications ranging from virtual in vivo histology to mapping fine subsurface vascular structures \cite{sun_lensless_2024}.

\subsection{Lensless 3D endoscopy through volumetric scattering in real biological tissue}
\label{subsec:biotissue}

\begin{figure}[t]
    \centering
    \includegraphics[width=\linewidth]{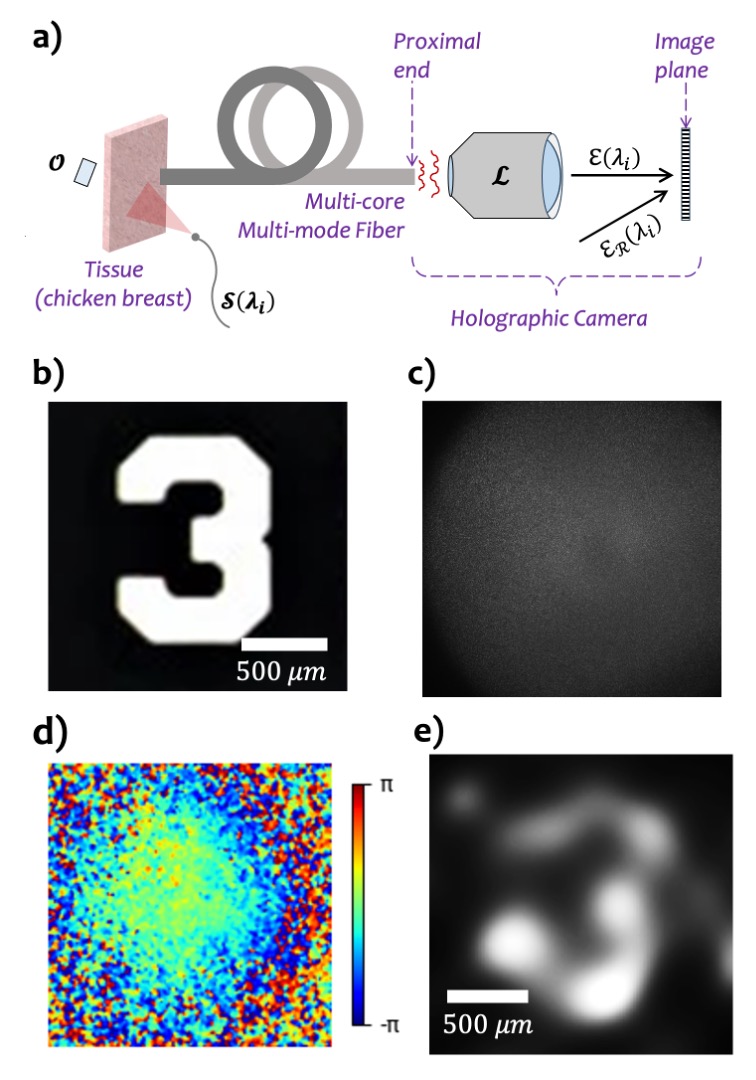}
    \caption{\textbf{Lensless endoscopy through a volumetric scattering medium.}\\
    a) Schematic illustration of the setup. A reflective test target placed behind a $5mm$ thick chicken breast sample is imaged using our method. b) Image of the obscured object c) Intensity image observed through the endoscope d) Phase map at the synthetic wavelength $\Lambda = 500~\mu m$ obtained using our approach and e) Reconstruction of the obscured object obtained by back propagating the field at the synthetic wavelength.}
    \label{fig:fig5}
\end{figure}

\noindent The experiments described above have been restricted to recovering holographic information of objects obscured by discrete scattering layers. In the following, we demonstrate the effectiveness of our approach through real biological tissue that exhibits strong volumetric scattering. 

In our experiment, schematically depicted in Fig.~\ref{fig:fig5}a,  we image a reflective target positioned behind a $5mm$ thick chicken breast sample. Unlike the static ground glass diffuser case described earlier, imaging through a real chicken breast sample is challenging due to multiple scattering and the dynamic fluctuations in speckle caused by small movements within the tissue. 
Since our approach makes no assumptions about the type of scattering, and is also robust to the temporal fluctuations due to our single-shot acquisition pipeline \cite{ballester_single-shot_2024}, we are able to successfully reconstruct a feature on the USAF target \textit{(Number 3 in Group 0, shown in Fig.~\ref{fig:fig5}b)} through the chicken breast sample in reflection geometry (Fig.~\ref{fig:fig5}e).

This experiment demonstrates the feasibility of our SWH based endoscope in more realistic biological media, highlighting its potential for use in in-vivo imaging applications, where multiple scattering and motion are prominent. Since our approach exploits the existence of spectral correlations, we anticipate that in more heavily scattering scenarios, we can still recover the information of the objects by increasing the synthetic wavelength, i.e., reducing the wavelength separation $\Delta\lambda$. This comes with the tradeoff of a further reduced spatial resolution. We discuss the tradespace between these factors in Supplementary Sec. 3.

\subsection{Scattering assisted Field-of-View (FoV) expansion}
\label{subsec:fovexpansion}
\begin{figure*}[t]
    \centering
    \includegraphics[width=0.8\linewidth]{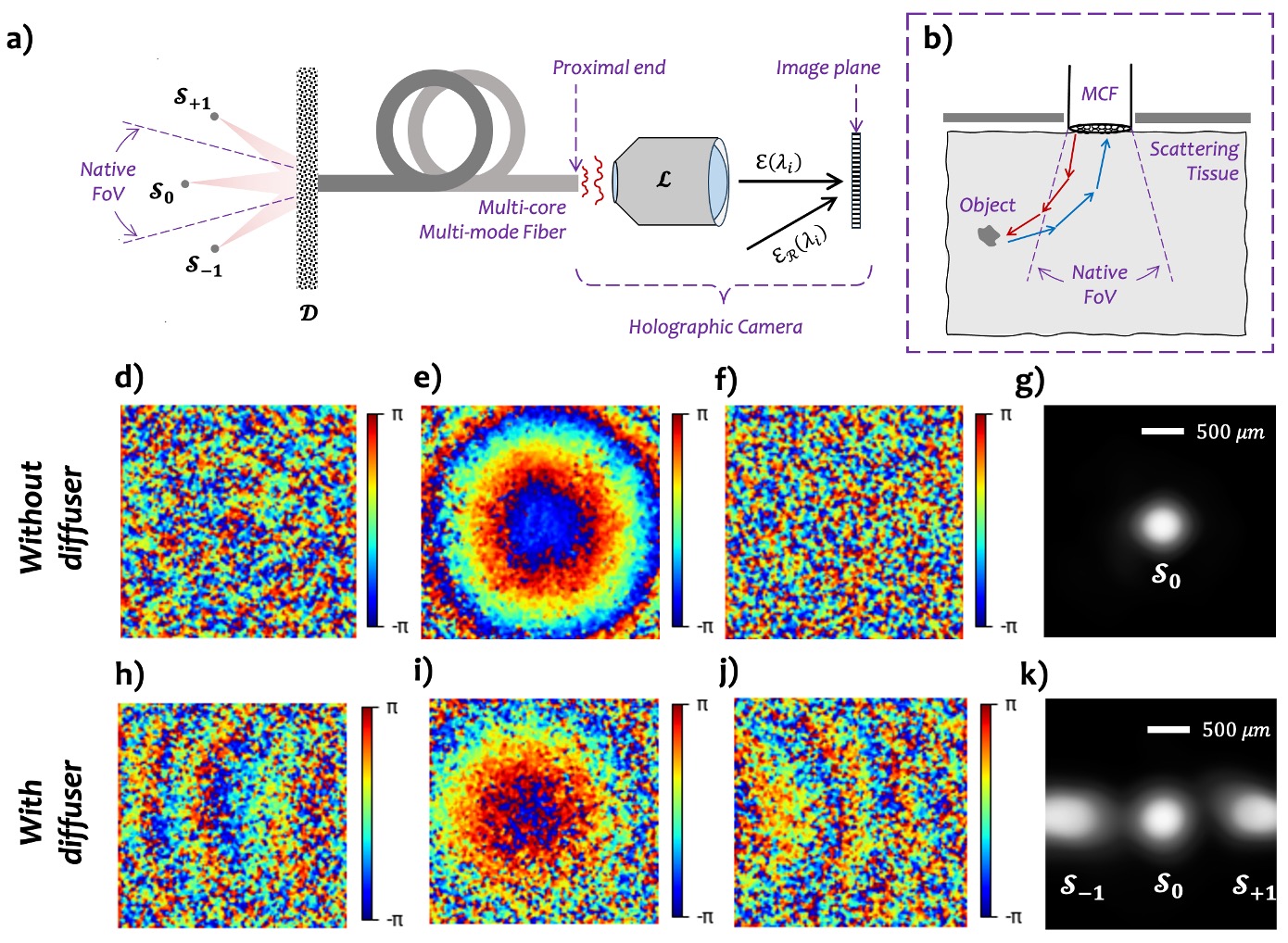}
    \caption{\textbf{Scattering assisted Field-of-View expansion.} a) \textit{Schematic of the experimental setup.} We acquire synthetic wavelength holograms of a point source by placing at three distinct angular positions behind a scattering medium. b) The scattering medium directs the incident light (red) to regions beyond the native FoV of the fiber. Similarly, light interacting with object beyond the native FoV of the fiber is also directed into the acceptance cone of the fiber (blue). d-f) Phase of the synthetic wavelength holograms from the three angular positions $\mathscr{S}_{-1}$, $\mathscr{S}_{0}$, and $\mathscr{S}_{+1}$ directly illuminating the M\textsuperscript{3}CF (without an intermediary diffuser, $\mathscr{D}$).g) Intensity of the reconstructed  points sources. h-j) Phase of the synthetic wavelength holograms from the three point sources $\mathscr{S}_{-1}$, $\mathscr{S}_{0}$, and $\mathscr{S}_{+1}$ in the presence of an intermediary diffuser, $\mathscr{D}$. k) Intensity of the reconstructed  points sources.}
    \label{fig:fig6}
\end{figure*}
\noindent The results presented in the previous sections demonstrate the ability of the SWH-based endoscope to recover three-dimensional information from object structures despite the presence of scattering. The ability to mitigate scattering effects at optical wavelengths opens new opportunities that are difficult to access with traditional optical endoscopes. In particular, the FoV of lensless endoscopes is typically restricted by the numerical aperture (NA) of the individual cores and the inter-core sampling spacing between the fibers in the MCF. This limits the probe volume to a narrow, “soda-straw-like” channel behind the distal end of the fiber. The wavefronts emerging from off-axis regions outside this narrow angular range are either aliased or fail to propagate through the MCF.

However, in the application scenario considered here, the presence of a scattering medium between the M\textsuperscript{3}CF and the object increases the étendue of the system~\cite{kuo2020high}. The wavefronts incident from the off-axis regions outside the original FoV can be scattered into the acceptance cone of the fiber, as illustrated in Fig.~\ref{fig:fig6}b. Thus, the scattering medium enables the MCF to collect light that is inaccessible to otherwise, and thereby extending the angular coverage of the endoscope. 

Although this étendue expansion can also occur in conventional optical endoscopes, the information collected from the extended FoV is typically scrambled by scattering and unusable. However, the ability of SWH to bypass scattering can be leveraged to unscramble the information within these scattered wavefronts and recover imagery over a significantly extended  FoV. Furthermore, if the inter-core spacing, $\delta x$ is smaller than $\frac{\Lambda}{2}$, our approach can theoretically image over the entire hemispherical FoV (see Supplementary Sec. 4), far surpassing the performance of traditional methods \cite{perperidis2020image}.

We show an experimental demonstration of this ability using the configuration depicted in Fig.~\ref{fig:fig6}a. Here, we illuminate the M\textsuperscript{3}CF by positioning a point source at three distinct angular positions: $\mathscr{S}_{-1}$, $\mathscr{S}_{0}$, and $\mathscr{S}_{+1}$, where $\mathscr{S}_{0}$ lies within the native FOV of the fiber, and $\mathscr{S}_{-1}$, and $\mathscr{S}_{+1}$ are on either side outside the native FoV. The phase maps in Fig.~\ref{fig:fig6}d--f show that, without a scattering medium, no synthetic-wavelength hologram is recovered for point sources at $\mathscr{S}{-1}$ and $\mathscr{S}{+1}$. This occurs because light from these off-axis positions are not coupled into the M\textsuperscript{3}CF. However, after scattering is introduced by placing a ground-glass diffuser ($\mathscr{D}$) between the M\textsuperscript{3}CF and the point source, holographic information can be captured in the form of spherical synthetic wavefronts emanating from point sources at $\mathscr{S}{-1}$, $\mathscr{S}{0}$, and $\mathscr{S}_{+1}$. We also reconstruct the intensity distribution for all three measurements by computationally backpropagating the hologram at the synthetic wavelength. An overlaid reconstruction for the three point sources can be seen in (Fig.~\ref{fig:fig6}k). Using this approach, we experimentally validated an expansion of the effective FOV from the original $23^\circ$ to approximately $46^\circ$ in our experiment, thus demonstrating nearly a twofold increase in angular coverage.

We note that our demonstrated FoV expansion technique stands in stark contrast to existing approaches that either rely on using additional optics or deep learning models trained on large datasets to learn complex mappings between scrambled and original wavefronts \cite{kuschmierz2021ultra,shanker_quantitative_2024,chen2025diffusion}. The latter, while effective within specific regimes, often lack generalizability and require extensive retraining for new samples or experimental configurations. In comparison, SWH offers a physics based, data independent framework for FoV expansion by leveraging scattering. We emphasize that this capability is enabled by the unique properties of synthetic waves that are generated using a carrier wave at optical wavelengths. More specifically, the scattering of the optical carrier waves ensures that the light is captured by the MCF, whereas, the assembled field at the longer synthetic wavelength provides the robustness to scattering, enabling the recovery of holographic information.

\subsection{Improved robustness to fiber bending}
\label{subsec:fiberbending}
\begin{figure*}[t]
    \centering
    \includegraphics[width=0.8\linewidth]{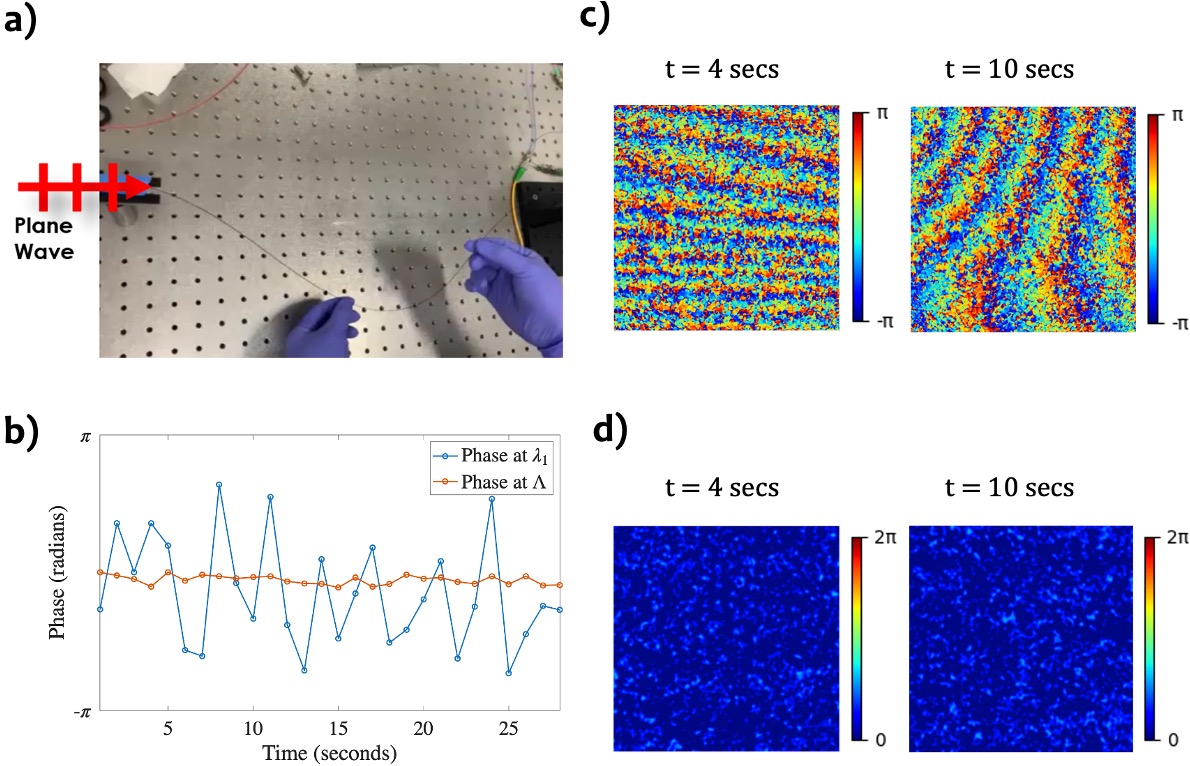}
    \caption{\textbf{Improved robustness to fiber bending.} a) We introduce path length perturbations at different time instances by bending the multicore fiber. b) The average phase over a pixel neighborhood (see text) plotted as a function of time. We see that the fluctuations in phase introduced by the bending of the fiber is significantly larger at the optical wavelength. c) Phase maps at the optical wavelengths. d) Phase maps at the synthetic wavelength at the same time instances. }
    \label{fig:fig7}
\end{figure*}

\noindent The use of synthetic waves also helps increase the robustness of fiber endoscopes against external perturbations such as fiber bending. This is primarily due to the fact that the path length variations introduced by these perturbations produce negligible phase variation at the longer synthetic wavelengths. We demonstrate this effect in the experiment shown in Fig.~\ref{fig:fig7}, where we illuminate the distal facet of a MCF (Sumida, FIGH-10-500N) using collimated beams from two lasers at $\lambda_1 \approx 854.33~nm$ and $\lambda_2 \approx 855.79~nm$. This results in a synthetic wavelength of $\Lambda \approx 500~\mu m$. We apply external perturbations by bending and twisting the MCF manually (see Fig.~\ref{fig:fig7}a). While applying these perturbations, we record a series of single-shot synthetic wavelength holograms at the proximal end of the fiber (Methods).

In Fig.~\ref{fig:fig7}c, we show the optical phase recorded at the proximal end of the fiber at two time instances while the fiber is bent and twisted. It can be seen that the perturbations introduced by the fiber bending, produce phase fluctuations at the optical wavelengths. However, the phase maps observed at the synthetic wavelength $\Lambda = 500~\mu m$ for the same time stamps (Fig.~\ref{fig:fig7}d), experience negligible fluctuations. To further quantify this effect, in Fig.~\ref{fig:fig7}b, we plot the temporal trace of the average phase over a $(11,11)$ pixel neighborhood in these two cases. We see that the optical phase fluctuations are significantly higher that the fluctuations at the synthetic wavelength, which largely remains constant. 

This observation confirms that the SWH-based endoscope maintains stability against the phase distortion caused by mechanical bending of the fiber. Furtheremore, this experiment validates the theoretical prediction that SWH provides intrinsic robustness to environmental perturbations. This robustness enables flexible endoscopic operation without the need for active phase stabilization through external phase modulators, thereby ensuring a stable performance under realistic operating conditions. 

\section{Discussion}
In this manuscript, we have demonstrated that SWH can be used to enhance imaging capabilities of fiber optic endoscopes. We experimentally recovered the images of objects hidden behind single scattering layers, as well as through  $5~mm$ thick biological volumetric scattering tissue with a spatial resolution of  $\approx 500~\mu m$. At the same time, we leverage the increased etendue of the imaging system in the presence of a scattering medium to expand the FoV of our endoscope. Using this, we demonstrated image recovery over an expanded FoV of $46\degree$ without needing any external modulators or distal optics.

SWH leverages the spectral correlations in the speckle fields recorded at closely spaced optical wavelengths, to recover the holographic information and thereby enable the capabilities demonstrated above. Unlike prior approaches that use spectral correlations for a variety of applications~\cite{french_snapshot_2018,groger2024two,metzger2017harnessing}, synthetic wavelength imaging is distinct in how these correlations are interpreted. In this paradigm, the beat wave between the two optical wavelengths is treated as an electromagnetic field at the synthetic wavelength~\cite{willomitzer_fast_2021,willomitzer2019synthetic,willomitzer2024synthetic}. This insight allows the resulting computational field to be analyzed using traditional computational optics methods, such as digital holography. In this work, we extended this concept to endoscopy and demonstrated the recovery of three-dimensional information through scattering.

Furthermore, unlike conventional wavefront shaping (WFS) techniques \cite{mosk2012controlling,cao2022shaping,cao2023controlling,rotter2017light}, which counteract the effects of disorder by physically controlling the spatial and temporal properties of light using spatial light modulators, our approach achieves a similar objective through computational means. Instead of relying on physical modulation of the illumination or real time feedback, we post-process scattered optical fields recorded at multiple wavelengths to computationally control light propagation in complex media.

However, SWH based endoscopy, introduced in this work, is not without limitations and tradeoffs. First, this study focused solely  on elastic scattering and reflectivity variations as the endogenous contrast mechanisms for image recovery. In future works, our  approach may also be extended to other contrast mechanisms, including absorption variations and temporal fluctuations in speckle caused by dynamic cellular activity, which could enable functional imaging. 

Furthermore, a key tradeoff of SWH in its current form is that the improved robustness to scattering comes at the expense of lower resolution in the reconstructed images \cite{willomitzer_fast_2021, willomitzer2024synthetic}. As discussed before, we reconstruct the hidden objects through computational back propagation of the assembled field at the synthetic wavelength. For this case, the lateral resolution is given by $\Lambda z / D$, consistent with the Rayleigh resolution criterion~\cite{goodman2008introduction}, where $\Lambda$ is the synthetic wavelength, $z$ is the distance from the object to the fiber tip, and $D$ is the aperture diameter. In our system, $D$ corresponds to the diameter of the M\textsuperscript{3}CF. In most experiments shown in this paper, we could realize a $\frac{D}{2z} \approx NA \approx 0.5$ resulting in a resolution of $\approx500 \mu m$ at a $\Lambda = 500 \mu m$. Further experiments verifying  that our system approaches this theoretical resolution bound can be found in Supplementary Sec.~2. 

Although this resolution is sufficient for potential applications such as visualizing larger tumor cell clusters or blood clots, several biomedical endoscopic imaging applications require higher resolution. A straightforward way to improve the resolution of our approach is to reduce the synthetic wavelength $\Lambda$. By increasing the wavelength separation, $\Delta\lambda$, we can effectively reduce the synthetic wavelength $\Lambda$, and thereby the spatial resolution. However, the minimal synthetic wavelength $\Lambda_{min}$, that can be realized in our approach is determined by the spectral memory effect of the speckle fields emerging from the multi-core fiber. As the wavelength separation increases, the path length error $\Psi$, introduced by the combination of the scattering medium and the modal dispersion through the fiber, exceeds the Rayleigh quarter wavelength criterion $\frac{\Lambda}{4}$ , causing the optical speckle fields to decorrelate \cite{willomitzer_fast_2021, willomitzer2024synthetic}. In the experiments shown here, we used multimode cores, that exhibit large modal dispersion and thereby limiting the largest wavelength separation, $\Delta\lambda$ to be $\approx 2~nm$, corresponding to a minimum achievable synthetic wavelength, $\Lambda_{\min} \approx 300\,\mathrm{\mu m}$. This value is consistent with the values quoted in literature for the same fiber \cite{french_snapshot_2018}. Additional details on the factors affecting the smallest achievable synthetic wavelength, $\Lambda_{\min}$, are provided in Supplementary Secs.~1 and 3.

In future work, we will explore strategies to address these limitations and improve the resolution of our technique. For instance, in the experiments presented herein, we used step-index multimode cores, which exhibit substantial modal dispersion and therefore constrain the minimum achievable synthetic wavelength. A more practical and scalable implementation could instead use graded-index multimode fibers, which can significantly minimize the modal dispersion \cite{rawson1980frequency}, and thereby enable the use of much smaller synthetic wavelengths. In addition, M\textsuperscript{3}CFs comprising single-mode or few-mode fiber cores can also be used to reduce the attainable synthetic wavelength and thereby improve spatial resolution. Further improvements in resolution may also be achievable from deep learning-based reconstruction methods \cite{chen2025diffusion} or synthetic aperture techniques \cite{song2024ptycho} that serve to increase the aperture diameter. Through these modifications, we anticipate that synthetic wavelengths as short as $50~\mu m$, comparable to those commonly used in 3D metrology can be achievable \cite{zhou2022review}. This strategy could provide a path toward translating our approach from applications with coarser resolution requirements to potential applications requiring finer resolution, such as optogenetics \cite{accanto2023flexible}.

\section{Methods}
\subsection{Single-shot off-axis holography acquisition scheme}
\label{subsec:offaxismethods}
\begin{figure*}[t]
    \centering
    \includegraphics[width=\linewidth]{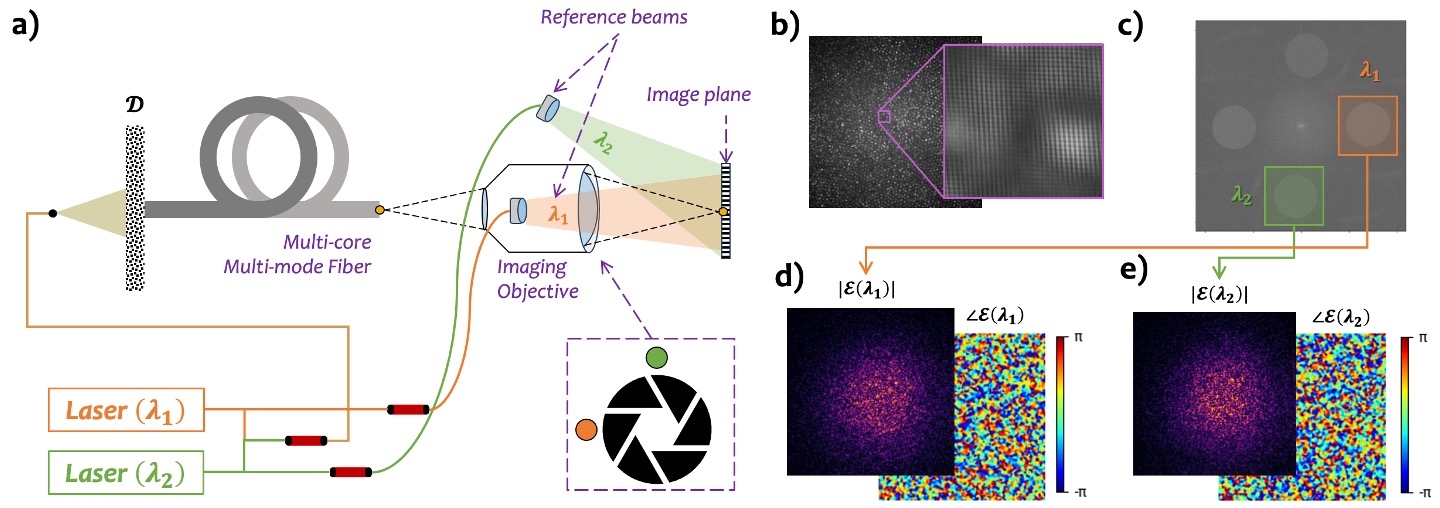}
    \caption{a)  \textit{Schematic of the single-shot synthetic wavelength hologram acquisition setup.} Each laser beam is split into an object beam and a reference beam. The object beams from the two lasers are combined together to illuminate the scattering medium. The two reference beams are oriented at orthogonal directions (see inset) to encode the optical fields at each wavelength in distinct bands in the Fourier domain. b) The single-shot hologram recorded at the image plane. Inset: a zoomed in view of the speckle cell overlaid with cross-hatched fringes from the two reference beams. c) The two-dimensional Fourier transform of the recorded single shot hologram where the distinct bands corresponding to the two wavelengths (arising from the orthogonal orientations of the reference beam) can be seen. d) and e) The amplitude and phase of the demodulated optical fields at wavelengths $\lambda_1$ and $\lambda_2$.}
    \label{fig:offaxisSetup}
\end{figure*}
To acquire the complex-valued optical fields at each wavelength, we rely on an off-axis holography scheme, shown in Fig.~\ref{fig:offaxisSetup}. Here, the laser emission is split into a probe beam and a reference beam. The probe beam illuminates the scattering medium, whereas the reference beam is collimated using a through a collimating lens and launched towards a CMOS FPA (Blackfly S Board-level USB3: 20 MP, Mono). The CMOS FPA we used has a pixel pitch of $2.4~\mu m$ and an array size of $5472 \times 3648$, spanning a total FoV of $13~mm \times 8.4~mm$. However, we crop the image to a central square region corresponding to $8.4~mm \times 8.4~mm$ area. We used a singlet lens with a focal length of $200~mm$, to image the distal end of an M\textsuperscript{3}CF (Edmund Optics, 304.8mm Standard Res Image Conduit with 50\(\mu\)m fiber cores) onto the FPA, with a magnification of $\approx 3\times$. At the center wavelength of $855 nm$ used in our system, the angular separation between the reference and the object beams can be as high as $10 \degree$, before the off-axis fringes can be aliased.

We used a narrow linewidth tunable laser (Toptica DFB) to source the probe and reference beams. The lasers output a maximum power of $140~mW$ and can be tuned over an $\approx 2~nm$ range from $854.2 ~nm$ to $856.6~nm$ by varying the operating temperature. We can realize a minimum synthetic wavelength of $0.3~mm$ using these lasers. For single-shot synthetic wavelength acquisition, we utilize the spatial multiplexing scheme illustrated in Fig.~\ref{fig:offaxisSetup}a. In this approach that our team has introduced in  \cite{ballester_single-shot_2024}, the object is illuminated simultaneously using probe beams sourced from two lasers operating at the desired wavelength separation. However, the reference beams from each laser are launched towards the FPA from orthogonal directions. This results in the generation of cross-hatched fringes, where the hologram at each wavelength is encoded around the carrier fringes that are oriented in orthogonal directions to each other.  However, this setup can also operate in a sequential-acquisition mode, in which the optical fields are captured as single-shot measurements in rapid succession by tuning the wavelength of one laser. Therefore, both implementations warrant the designation “single-shot setup.”

To isolate the holograms at each optical wavelength, we compute the two-dimensional Fourier transform of the acquired intensity image. As shown in Fig.~\ref{fig:offaxisSetup}c, the complex-valued field information at each wavelength is centered around a band of frequencies around the carrier frequency. These bands are filtered, and demodulated in the Fourier domain, and then inverse Fourier transformed to recover the complex-valued optical field at each wavelength. We backpropagate this field at the corresponding synthetic wavelength $\Lambda$ using the angular spectrum method. 

\subsection{Experimental setup for Lensless 3D endoscopy through scattering layers}
\label{subsec:scatexpsetup}

\noindent We use the apparatus shown in Fig.~\ref{fig:fig2} to demonstrate the operation of SWH based lensless endoscopy through  scattering layers in a transmissive configuration. In this experiment, we recover holographic information of a transmissive USAF target, $\mathscr{O}$, embedded between two ground glass diffusers, $\mathscr{D}_{1}$ and $\mathscr{D}_{2}$, both 1500 grit. The illumination beam from the tunable laser was expanded to 10 mm diameter to illuminate $\mathscr{D}_{1}$. The scattered light illuminates a small region corresponding to Group $0$, Element $5$ on the target $\mathscr{O}$ (Fig.~\ref{fig:fig2}b), placed $5~mm$ away from $\mathscr{D}_{1}$. The light from the USAF target then propagates $5~mm$ through free space to reach $\mathscr{D}_{2}$. The light is coupled into the M\textsuperscript{3}CF placed close to $\mathscr{D}_{2}$. The proximal end of the M\textsuperscript{3}CF is imaged onto the sensor. In the results shown in Fig~\ref{fig:fig2}c,d, the two lasers were operated at $\lambda_1 \approx 854.33~nm$ and $\lambda_2 \approx 855.79~nm$, resulting in a synthetic wavelength of $\Lambda \approx 500~\mu m$.

For the experiments in reflective configuration shown in Fig.~\ref{fig:fig3}, we used the mask of the letter "A" struck to a mirror as the obscured object, $\mathscr{O}$. The object was placed behind the ground glass diffuser $\mathscr{D}$ (1500 grit). As earlier, the illumination beam was expanded to 10 mm diameter to illuminate $\mathscr{D}$ at an incident angle of $30\degree$ (Fig.~\ref{fig:fig3}a). The scattered light illuminates $\mathscr{O}$ that is placed $\approx 5~mm$ away behind $\mathscr{D}$. The light reflected from the target then propagates $\approx 5~mm$ back to reach a different region on $\mathscr{D}$. The light coupled into the M\textsuperscript{3}CF placed close to $\mathscr{D}$ is imaged onto the sensor. In the results shown in Fig~\ref{fig:fig3}c,d, the two lasers were operated at $\lambda_1 \approx 854.33~nm$ and $\lambda_2 \approx 855.79~nm$, resulting in a synthetic wavelength of $\Lambda \approx 500~\mu m$. 
\subsection{Volumetric 3D reconstruction from fields assembled at multiple synthetic wavelengths}
\label{subsec:slifexpsetup}

\noindent For the volumetric reconstructions shown in Fig.~\ref{fig:fig4}, we acquire the complex-valued optical fields at a series of equally spaced in optical frequency $\mathscr{E} (\lambda_1), \mathscr{E} (\lambda_2), ...\mathscr{E} (\lambda_{N+1})$. From these measurements, we assemble a series of synthetic wavelength holograms, $\mathscr{E} (\Lambda_1), \mathscr{E} (\Lambda_2), ...\mathscr{E} (\Lambda_N)$ with a  synthetic wavelength $\Lambda_m = \frac{\lambda_{m+1}\lambda_1}{|\lambda_{m+1} - \lambda_1|}$. The resulting synthetic pulse front, shown in Fig.~\ref{fig:fig4}, is assembled by assigning the appropriate temporal phase to each synthetic wavelength field and then coherently summing the resulting fields, i.e.,
\[
\mathscr{P}(x,y,t)
=
\sum_{n=1}^{N}
\mathscr{E}(x,y,\Lambda_n)e^{\left(j\frac{2\pi c}{\Lambda_n}t\right)}
\]

In the experiments shown in Fig.~\ref{fig:fig4}a, we placed two reflective pins, with a pinhead diameter of $0.8~mm$ at two distinct axial positions separated by $1.5~mm$ behind a ground glass diffuser. We illuminate the  diffuser, $\mathscr{D}$, and acquired $41$ optical fields at series of wavelengths from $855~nm$ to $856~nm$ in steps of $\approx 0.1~nm$. From these, we assembled a series of synthetic wavelength holograms as described above. 

The width of the pulse assembled from these synthetic wavelength holograms is can be obtained as $\frac{\lambda^2}{c\Delta\lambda}$. In our case, this corresponds to a pulse width of $1.67~ps$, and a corresponding depth resolution of $500~\mu m$. In Fig.~\ref{fig:fig4}b-c, we show the reconstructed intensity at $z = 1~mm$ and $z = 1.5~mm$, where we can see that the object are distinctly separated axially. This is also apparent in the three-dimensional point cloud visualization shown in Fig.~\ref{fig:fig4}d. Additional details of this approach can be found in \cite{cornwall2024synthetic}. \\

\subsection{Data availability}
The raw data may be obtained by contacting the authors.

\subsection{Code availability}
The code may be obtained by contacting the authors.

\section{References}

\bibliography{bibliography}

@article{french_snapshot_2018,
	title = {Snapshot fiber spectral imaging using speckle correlations and compressive sensing},
	volume = {26},
	issn = {1094-4087},
	url = {https://opg.optica.org/oe/abstract.cfm?uri=oe-26-24-32302},
	doi = {10.1364/OE.26.032302},
	language = {EN},
	number = {24},
	urldate = {2025-09-29},
	journal = {Optics Express},
	author = {French, Rebecca and Gigan, Sylvain and Muskens, Otto l},
	month = nov,
	year = {2018},
	note = {Publisher: Optica Publishing Group},
	pages = {32302--32316},
}

@article{sun_quantitative_2022,
	title = {Quantitative phase imaging through an ultra-thin lensless fiber endoscope},
	volume = {11},
	copyright = {2022 The Author(s)},
	issn = {2047-7538},
	url = {https://www.nature.com/articles/s41377-022-00898-2},
	doi = {10.1038/s41377-022-00898-2},
	language = {en},
	number = {1},
	urldate = {2025-09-29},
	journal = {Light: Science \& Applications},
	author = {Sun, Jiawei and Wu, Jiachen and Wu, Song and Goswami, Ruchi and Girardo, Salvatore and Cao, Liangcai and Guck, Jochen and Koukourakis, Nektarios and Czarske, Juergen W.},
	month = jul,
	year = {2022},
	note = {Publisher: Nature Publishing Group},
	pages = {204},
}

@article{sun_lensless_2024,
	title = {Lensless fiber endomicroscopy in biomedicine},
	volume = {5},
	copyright = {2024 The Author(s)},
	issn = {2662-1991},
	url = {https://link-springer-com.ezproxy2.library.arizona.edu/article/10.1186/s43074-024-00133-8},
	doi = {10.1186/s43074-024-00133-8},
	language = {en},
	number = {1},
	urldate = {2025-09-29},
	journal = {PhotoniX},
	author = {Sun, Jiawei and Kuschmierz, Robert and Katz, Ori and Koukourakis, Nektarios and Czarske, Juergen W.},
	month = dec,
	year = {2024},
	note = {Number: 1
Publisher: SpringerOpen},
	pages = {1--9},
}

@article{badt_real-time_2022,
	title = {Real-time holographic lensless micro-endoscopy through flexible fibers via fiber bundle distal holography},
	volume = {13},
	copyright = {2022 The Author(s)},
	issn = {2041-1723},
	url = {https://www.nature.com/articles/s41467-022-33462-y},
	doi = {10.1038/s41467-022-33462-y},
	language = {en},
	number = {1},
	urldate = {2025-09-29},
	journal = {Nature Communications},
	author = {Badt, Noam and Katz, Ori},
	month = oct,
	year = {2022},
	note = {Publisher: Nature Publishing Group},
	pages = {6055},
}

@article{willomitzer_fast_2021,
	title = {Fast non-line-of-sight imaging with high-resolution and wide field of view using synthetic wavelength holography},
	volume = {12},
	copyright = {2021 The Author(s)},
	issn = {2041-1723},
	url = {https://www.nature.com/articles/s41467-021-26776-w},
	doi = {10.1038/s41467-021-26776-w},
	language = {en},
	number = {1},
	urldate = {2025-09-29},
	journal = {Nature Communications},
	author = {Willomitzer, Florian and Rangarajan, Prasanna V. and Li, Fengqiang and Balaji, Muralidhar M. and Christensen, Marc P. and Cossairt, Oliver},
	month = nov,
	year = {2021},
	pages = {6647},
}

@article{ballester_single-shot_2024,
	title = {Single-shot synthetic wavelength imaging: {Sub}-mm precision {ToF} sensing with conventional {CMOS} sensors},
	volume = {178},
	issn = {0143-8166},
	shorttitle = {Single-shot synthetic wavelength imaging},
	url = {https://www.sciencedirect.com/science/article/pii/S0143816624001441},
	doi = {10.1016/j.optlaseng.2024.108165},
	urldate = {2025-09-29},
	journal = {Optics and Lasers in Engineering},
	author = {Ballester, Manuel and Wang, Heming and Li, Jiren and Cossairt, Oliver and Willomitzer, Florian},
	month = jul,
	year = {2024},
	pages = {108165},
}

@article{porat_widefield_2016,
	title = {Widefield lensless imaging through a fiber bundle via speckle correlations},
	volume = {24},
	copyright = {© 2016 Optical Society of America},
	issn = {1094-4087},
	url = {https://opg.optica.org/oe/abstract.cfm?uri=oe-24-15-16835},
	doi = {10.1364/OE.24.016835},
	language = {EN},
	number = {15},
	urldate = {2025-09-29},
	journal = {Optics Express},
	author = {Porat, Amir and Andresen, Esben Ravn and Rigneault, Hervé and Oron, Dan and Gigan, Sylvain and Katz, Ori},
	month = jul,
	year = {2016},
	note = {Publisher: Optica Publishing Group},
	pages = {16835--16855},
}

@article{shanker_quantitative_2024,
	title = {Quantitative phase imaging endoscopy with a metalens},
	volume = {13},
	copyright = {2024 The Author(s)},
	issn = {2047-7538},
	url = {https://www.nature.com/articles/s41377-024-01587-y},
	doi = {10.1038/s41377-024-01587-y},
	language = {en},
	number = {1},
	urldate = {2025-09-29},
	journal = {Light: Science \& Applications},
	author = {Shanker, Aamod and Fröch, Johannes E. and Mukherjee, Saswata and Zhelyeznyakov, Maksym and Brunton, Steven L. and Seibel, Eric J. and Majumdar, Arka},
	month = nov,
	year = {2024},
	note = {Publisher: Nature Publishing Group},
	pages = {305},
}

@article{willomitzer2019synthetic,
  title={Synthetic wavelength holography: An extension of Gabor's holographic principle to imaging with scattered wavefronts},
  author={Willomitzer, Florian and Rangarajan, Prasanna V and Li, Fengqiang and Balaji, Muralidhar M and Christensen, Marc P and Cossairt, Oliver},
  journal={arXiv preprint arXiv:1912.11438},
  year={2019}
}

@article{willomitzer2024synthetic,
  title={Synthetic wavelength imaging: Utilizing spectral correlations for high-precision time-of-flight sensing},
  author={Willomitzer, Florian},
  journal={Computational Imaging for Scene Understanding: Transient, Spectral, and Polarimetric Analysis},
  volume={187},
  year={2024},
  publisher={John Wiley \& Sons}
}

@article{metzger2017harnessing,
  title={Harnessing speckle for a sub-femtometre resolved broadband wavemeter and laser stabilization},
  author={Metzger, Nikolaus Klaus and Spesyvtsev, Roman and Bruce, Graham D and Miller, Bill and Maker, Gareth T and Malcolm, Graeme and Mazilu, Michael and Dholakia, Kishan},
  journal={Nature communications},
  volume={8},
  number={1},
  pages={15610},
  year={2017},
  publisher={Nature Publishing Group UK London}
}

@article{mosk2012controlling,
  title={Controlling waves in space and time for imaging and focusing in complex media},
  author={Mosk, Allard P and Lagendijk, Ad and Lerosey, Geoffroy and Fink, Mathias},
  journal={Nature photonics},
  volume={6},
  number={5},
  pages={283--292},
  year={2012},
  publisher={Nature Publishing Group UK London}
}

@article{cao2022shaping,
  title={Shaping the propagation of light in complex media},
  author={Cao, Hui and Mosk, Allard Pieter and Rotter, Stefan},
  journal={Nature Physics},
  volume={18},
  number={9},
  pages={994--1007},
  year={2022},
  publisher={Nature Publishing Group UK London}
}

@article{cao2023controlling,
  title={Controlling light propagation in multimode fibers for imaging, spectroscopy, and beyond},
  author={Cao, Hui and {\v{C}}i{\v{z}}m{\'a}r, Tom{\'a}{\v{s}} and Turtaev, Sergey and Tyc, Tom{\'a}{\v{s}} and Rotter, Stefan},
  journal={Advances in Optics and Photonics},
  volume={15},
  number={2},
  pages={524--612},
  year={2023},
  publisher={Optica Publishing Group}
}

@article{rotter2017light,
  title={Light fields in complex media: Mesoscopic scattering meets wave control},
  author={Rotter, Stefan and Gigan, Sylvain},
  journal={Reviews of Modern Physics},
  volume={89},
  number={1},
  pages={015005},
  year={2017},
  publisher={APS}
}

@article{perperidis2020image,
  title={Image computing for fibre-bundle endomicroscopy: A review},
  author={Perperidis, Antonios and Dhaliwal, Kevin and McLaughlin, Stephen and Vercauteren, Tom},
  journal={Medical image analysis},
  volume={62},
  pages={101620},
  year={2020},
  publisher={Elsevier}
}

@article{bae2023feasibility,
  title={Feasibility studies of multimodal nonlinear endoscopy using multicore fiber bundles for remote scanning from tissue sections to bulk organs},
  author={Bae, Hyeonsoo and Rodewald, Marko and Meyer-Zedler, Tobias and Bocklitz, Thomas W and Matz, Gregor and Messerschmidt, Bernhard and Press, Adrian T and Bauer, Michael and Guntinas-Lichius, Orlando and Stallmach, Andreas and others},
  journal={Scientific reports},
  volume={13},
  number={1},
  pages={13779},
  year={2023},
  publisher={Nature Publishing Group UK London}
}

@article{conkey2016lensless,
  title={Lensless two-photon imaging through a multicore fiber with coherence-gated digital phase conjugation},
  author={Conkey, Donald B and Stasio, Nicolino and Morales-Delgado, Edgar E and Romito, Marilisa and Moser, Christophe and Psaltis, Demetri},
  journal={Journal of biomedical optics},
  volume={21},
  number={4},
  pages={045002--045002},
  year={2016},
  publisher={Society of Photo-Optical Instrumentation Engineers}
}

@article{gau2024multicore,
  title={Multicore fiber optic imaging reveals that astrocyte calcium activity in the mouse cerebral cortex is modulated by internal motivational state},
  author={Gau, Yung-Tian A and Hsu, Eric T and Cha, Richard J and Pak, Rebecca W and Looger, Loren L and Kang, Jin U and Bergles, Dwight E},
  journal={Nature communications},
  volume={15},
  number={1},
  pages={3039},
  year={2024},
  publisher={Nature Publishing Group UK London}
}

@article{richardson2013space,
  title={Space-division multiplexing in optical fibres},
  author={Richardson, David J and Fini, John M and Nelson, Lynn E},
  journal={Nature photonics},
  volume={7},
  number={5},
  pages={354--362},
  year={2013},
  publisher={Nature Publishing Group UK London}
}

@article{potter2024clinical,
  title={Clinical and biomedical applications of lensless holographic microscopy},
  author={Potter, Colin J and Xiong, Zhen and McLeod, Euan},
  journal={Laser \& Photonics Reviews},
  volume={18},
  number={10},
  pages={2400197},
  year={2024},
  publisher={Wiley Online Library}
}

@article{choi2022flexible,
  title={Flexible-type ultrathin holographic endoscope for microscopic imaging of unstained biological tissues},
  author={Choi, Wonjun and Kang, Munkyu and Hong, Jin Hee and Katz, Ori and Lee, Byunghak and Kim, Guang Hoon and Choi, Youngwoon and Choi, Wonshik},
  journal={Nature communications},
  volume={13},
  number={1},
  pages={4469},
  year={2022},
  publisher={Nature Publishing Group UK London}
}

@article{dremel2026lensless,
  title={Lensless single-shot multicore fiber endomicroscopy using a single multispectral hologram},
  author={Dremel, Jakob and Scharf, Elias and Richter, Sven and Czarske, J{\"u}rgen and Kuschmierz, Robert},
  journal={Light: Advanced Manufacturing},
  volume={6},
  number={4},
  pages={896--903},
  year={2026},
  publisher={Light: Advanced Manufacturing}
}

@article{scharf2019holographic,
  title={Holographic lensless fiber endoscope with needle size using self-calibration},
  author={Scharf, Elias and Kuschmierz, Robert and Czarske, J{\"u}rgen},
  journal={tm-Technisches Messen},
  volume={86},
  number={3},
  pages={144--150},
  year={2019},
  publisher={De Gruyter Oldenbourg}
}

@article{du2022hybrid,
  title={Hybrid multimode-multicore fibre based holographic endoscope for deep-tissue neurophotonics},
  author={Du, Yang and Turtaev, Sergey and Leite, Ivo T and Lorenz, Adrian and Kobelke, Jens and Wondraczek, Katrin and {\v{C}}i{\v{z}}m{\'a}r, Tom{\'a}{\v{s}}},
  journal={Light: Advanced Manufacturing},
  volume={3},
  number={3},
  pages={408--416},
  year={2022},
  publisher={Light: Advanced Manufacturing}
}

@article{song2024ptycho,
  title={Ptycho-endoscopy on a lensless ultrathin fiber bundle tip},
  author={Song, Pengming and Wang, Ruihai and Loetgering, Lars and Liu, Jia and Vouras, Peter and Lee, Yujin and Jiang, Shaowei and Feng, Bin and Maiden, Andrew and Yang, Changhuei and others},
  journal={Light: Science \& Applications},
  volume={13},
  number={1},
  pages={168},
  year={2024},
  publisher={Nature Publishing Group UK London}
}

@article{zhou2022review,
  title={A review of the dual-wavelength technique for phase imaging and 3D topography},
  author={Zhou, Haowen and Hussain, Mallik MR and Banerjee, Partha P},
  journal={Light: Advanced Manufacturing},
  volume={3},
  number={2},
  pages={314--334},
  year={2022},
  publisher={Light: Advanced Manufacturing}
}

@article{chen2025diffusion,
  title={Diffusion-driven lensless fiber endomicroscopic quantitative phase imaging towards digital pathology},
  author={Chen, Zhaoging and Sun, Jiawei and Yang, Xibin and Ye, Xinyi and Zhao, Bin and Li, Xuelong and Czarske, Juergen W and others},
  journal={Advanced Imaging},
  volume={2},
  number={4},
  pages={041003},
  year={2025},
  publisher={Editorial Office of Advanced Imaging}
}

@article{goodman2008introduction,
  title={Introduction to Fourier optics. 2005},
  author={Goodman, Joseph W},
  journal={Roberts and Company},
  year={2008}
}

@article{rawson1980frequency,
  title={Frequency dependence of modal noise in multimode optical fibers},
  author={Rawson, Eric G and Goodman, Joseph W and Norton, Robert E},
  journal={Journal of the Optical Society of America},
  volume={70},
  number={8},
  pages={968--976},
  year={1980},
  publisher={Optical Society of America}
}

@article{cornwall2024synthetic,
  title={Synthetic Light-in-Flight},
  author={Cornwall, Patrick and Ballester, Manuel and Forschner, Stefan and Madabhushi Balaji, Muralidhar and Katsaggelos, Aggelos and Willomitzer, Florian},
  journal={arXiv e-prints},
  pages={arXiv--2407},
  year={2024}
}

@article{accanto2023flexible,
  title={A flexible two-photon fiberscope for fast activity imaging and precise optogenetic photostimulation of neurons in freely moving mice},
  author={Accanto, Nicol{\`o} and Blot, Fran{\c{c}}ois GC and Lorca-C{\'a}mara, Antonio and Zampini, Valeria and Bui, Florence and Tourain, Christophe and Badt, Noam and Katz, Ori and Emiliani, Valentina},
  journal={Neuron},
  volume={111},
  number={2},
  pages={176--189},
  year={2023},
  publisher={Elsevier}
}

@article{groger2024two,
  title={Two-wavelength holographic micro-endoscopy},
  author={Gr{\"o}ger, Alexander and Kuschmierz, Robert and Birk, Alexander and Pedrini, Giancarlo and Reichelt, Stephan},
  journal={Optics Express},
  volume={32},
  number={13},
  pages={23687--23701},
  year={2024},
  publisher={Optica Publishing Group}
}

@article{kuschmierz2021ultra,
  title={Ultra-thin 3D lensless fiber endoscopy using diffractive optical elements and deep neural networks},
  author={Kuschmierz, Robert and Scharf, Elias and Orteg{\'o}n-Gonz{\'a}lez, David F and Glosemeyer, Tom and Czarske, J{\"u}rgen W},
  journal={Light: Advanced Manufacturing},
  volume={2},
  number={4},
  pages={415--424},
  year={2021},
  publisher={Light: Advanced Manufacturing}
}

@inproceedings{forschner2024towards,
  title={Towards synthetic wavelength imaging through multi-mode fibers},
  author={Forschner, Stefan and Cornwall, Patrick and Ballester, Manuel and Madabhushi Balaji, Muralidhar and Czarske, J{\"u}rgen and Willomitzer, Florian},
  booktitle={Proc. of SPIE Vol},
  volume={13258},
  pages={1325801--212},
  year={2024}
}

@inproceedings{balaji2025fiber,
  title={Fiber endoscopy using synthetic wavelengths for 3D tissue imaging},
  author={Balaji, Muralidhar Madabhushi and Cornwall, Patrick and Liu, Parker and Forschner, Stefan and Czarske, J{\"u}rgen and Willomitzer, Florian},
  booktitle={Computational Optical Imaging and Artificial Intelligence in Biomedical Sciences II},
  volume={13333},
  pages={18--22},
  year={2025},
  organization={SPIE}
}

@inproceedings{liu20263d,
  title={3D endoscopy in scattering environments using synthetic wavelength holography},
  author={Liu, Pengyu and Balaji, Muralidhar Madabhushi and Cornwall, Patrick and Czarske, J{\"u}rgen and Willomitzer, Florian},
  booktitle={Computational Optical Imaging and Artificial Intelligence in Biomedical Sciences III},
  volume={13865},
  pages={14--17},
  year={2026},
  organization={SPIE}
}

@article{kuo2020high,
  title={High resolution {\'e}tendue expansion for holographic displays},
  author={Kuo, Grace and Waller, Laura and Ng, Ren and Maimone, Andrew},
  journal={ACM Transactions on Graphics (TOG)},
  volume={39},
  number={4},
  pages={66--1},
  year={2020},
  publisher={ACM New York, NY, USA}
}

@article{kassem2025intensity,
  title={Intensity-Correlation Synthetic Wavelength Imaging in Dynamic Scattering Media},
  author={Kassem, Khaled and Fatima, Areeba and Cornwall, Patrick and Balaji, Muralidhar Madabhushi and Faccio, Daniele and Willomitzer, Florian},
  journal={arXiv preprint arXiv:2510.27620},
  year={2025}
}

@article{balaji2024probing,
  title={Probing diffusive media through speckle differencing},
  author={Balaji, Muralidhar Madabhushi and Ahsanullah, Danyal and Rangarajan, Prasanna},
  journal={Biomedical Optics Express},
  volume={15},
  number={9},
  pages={5442--5460},
  year={2024},
  publisher={Optica Publishing Group}
}

@article{katz2014non,
  title={Non-invasive single-shot imaging through scattering layers and around corners via speckle correlations},
  author={Katz, Ori and Heidmann, Pierre and Fink, Mathias and Gigan, Sylvain},
  journal={Nature photonics},
  volume={8},
  number={10},
  pages={784--790},
  year={2014},
  publisher={Nature Publishing Group}
}

@article{edrei2016optical,
  title={Optical imaging through dynamic turbid media using the Fourier-domain shower-curtain effect},
  author={Edrei, Eitan and Scarcelli, Giuliano},
  journal={Optica},
  volume={3},
  number={1},
  pages={71--74},
  year={2016},
  publisher={Optical Society of America}
}

@article{lindell2020three,
  title={Three-dimensional imaging through scattering media based on confocal diffuse tomography},
  author={Lindell, David B and Wetzstein, Gordon},
  journal={Nature communications},
  volume={11},
  number={1},
  pages={4517},
  year={2020},
  publisher={Nature Publishing Group UK London}
}

@article{bertolotti2012non,
  title={Non-invasive imaging through opaque scattering layers},
  author={Bertolotti, Jacopo and Van Putten, Elbert G and Blum, Christian and Lagendijk, Ad and Vos, Willem L and Mosk, Allard P},
  journal={Nature},
  volume={491},
  number={7423},
  pages={232--234},
  year={2012},
  publisher={Nature Publishing Group UK London}
}

@article{yoon2020deep,
  title={Deep optical imaging within complex scattering media},
  author={Yoon, Seokchan and Kim, Moonseok and Jang, Mooseok and Choi, Youngwoon and Choi, Wonjun and Kang, Sungsam and Choi, Wonshik},
  journal={Nature Reviews Physics},
  volume={2},
  number={3},
  pages={141--158},
  year={2020},
  publisher={Nature Publishing Group UK London}
}

@article{bertolotti2022imaging,
  title={Imaging in complex media},
  author={Bertolotti, Jacopo and Katz, Ori},
  journal={Nature Physics},
  volume={18},
  number={9},
  pages={1008--1017},
  year={2022},
  publisher={Nature Publishing Group UK London}
}

@article{feng2023neuws,
  title={NeuWS: Neural wavefront shaping for guidestar-free imaging through static and dynamic scattering media},
  author={Feng, Brandon Y and Guo, Haiyun and Xie, Mingyang and Boominathan, Vivek and Sharma, Manoj K and Veeraraghavan, Ashok and Metzler, Christopher A},
  journal={Science Advances},
  volume={9},
  number={26},
  pages={eadg4671},
  year={2023},
  publisher={American Association for the Advancement of Science}
}

@inproceedings{rangarajan2019spatially,
  title={Spatially resolved indirect imaging of objects beyond the line of sight},
  author={Rangarajan, Prasanna and Willomitzer, Florian and Cossairt, Oliver and Christensen, Marc P},
  booktitle={Unconventional and Indirect Imaging, Image Reconstruction, and Wavefront Sensing 2019},
  volume={11135},
  pages={124--131},
  year={2019},
  organization={SPIE}
}

\end{document}